\documentclass[%
 reprint, 
 amsmath,amssymb, 
 aps,prl,
]{revtex4-1}

\usepackage{mathtools} 
\usepackage{multirow} 
\usepackage{color} 
\usepackage{epsf} 
\usepackage{soul}
\usepackage{epsfig} 
\usepackage{ulem} 
\usepackage{graphicx}
\usepackage{dcolumn}
\usepackage{bm}
 
\begin{document}

\title{Uniaxial pressure induced half-metallic ferromagnetic phase transition in LaMnO$_3$}  

\author{Pablo Rivero$^1$, Vincent Meunier$^2$, and William Shelton$^{1}$} 
 
\email{jprivero@lsu.edu}

\affiliation{1. Center for Computation and Technology, Louisiana State University, Baton Rouge, Louisiana 70803, USA \\ 
2. Department of Physics, Applied Physics, and Astronomy, Rensselaer Polytechnic Institute, Troy, NY 12180, USA }

\date{\today}
 
\begin{abstract} 

 We use first-principles theory to predict that the application of uniaxial compressive strain leads to a transition from an antiferromagnetic insulator to a ferromagnetic half-metal phase in LaMnO$_3$. We identify the Q2 Jahn-Teller mode as the primary mechanism that drives the transition, indicating that this mode can be used to tune the lattice, charge, and spin coupling. Applying $\simeq$ 6 GPa of uniaxial pressure along the [010] direction activates the transition to a half-metallic $\textit{pseudo-cubic}$ state. The half-metallicity opens the possibility of producing colossal magnetoresistance in the stoichiometric LaMnO$_3$ compound at significantly lower pressure compared to recently observed investigations using hydrostatic pressure.

\begin{description} 
\item[PACS numbers] 
71.30.+h, 31.15.A-, 81.40.Vw, 71.70.Ej, 75.47.Gk 
 
\end{description} 
\end{abstract} 
 
\pacs{Valid PACS appear here}
\maketitle 
 
 
\section{Introduction} 
Lanthanum manganite (LaMnO$_3$) is a distorted perovskite material whose electronic and magnetic properties have potential technological use in a variety of applications such as pseudo-capacitors, batteries, and spintronics \cite{pseudocapacitors,batteries,bias,spintronic1,spintronic2}. Electron-electron correlation and lattice distortion effects are among the main drivers behind the appearance of these properties \cite{kanamori}, which depend on a number of parameters such as pressure, doping, and temperature. Of particular interest is the onset of colossal magnetoresistance (CMR) that can appear upon application of a large hydrostatic pressure \cite{hydrostatic}. This pressure induces a phase separated (PS) state characterized by the coexistence of different magnetic and structural domains. The PS state along with half-metallic character have been suggested by several authors to be key for optimizing the CMR state in manganites \cite{percolation,percolation2,mixed1,mixed2,dagotto,book,hf1,hf2}.

CMR is a phenomena characterized by large changes in electrical resistance in the presence of a magnetic field \cite{jonker,CMR1}. Hole-doped LaMnO$_3$ compounds (La$_{1-x}$A$_x$MnO$_3$, where A = Ca, Sr, Ba) feature the largest reported CMR effect, which is several orders of magnitude larger than that of other systems \cite{doped1,doped2,CMR2}. In these materials, CMR can be observed either at the phase transition from a ferromagnetic (FM) half-metallic state to paramagnetic (PM) insulating state or in the melting of the charge-orbital ordered state \cite{CO1,CO2}. The recent discovery of CMR in the stoichiometric LaMnO$_3$ under 32 GPa of hydrostatic pressure \cite{hydrostatic} demonstrates the possibility of obtaining this effect in the ordered LaMnO$_3$ compound. However, the CMR effect reported in this experiment was significantly reduced, probably because the system was not in a half-metallic state. The half-metallic domains necessary for achieving the optimal effect are predicted to occur at very high hydrostatic pressures, in excess of 100 GPa \cite{franchini2}, clearly showing the need to devise alternative strategies that would require more manageable applied strains. 

In this paper, we show that the pressure needed to obtain a FM half-metallic state is reduced by an order of magnitude when the pressure is applied along a single axis (uniaxial). We also show that the mechanism associated with the insulator-to-metal (IMT), structural, and magnetic phase transitions observed in this system are mainly controlled by a particular Jahn-Teller (JT) mode, as previously highlighted \cite{our,other1,other2}. This identification of a particular phonon-type mode, provides a clear strategy for tuning properties in other systems where such structural distortions exist.

\begin{figure}  
\includegraphics[width=0.48 \textwidth]{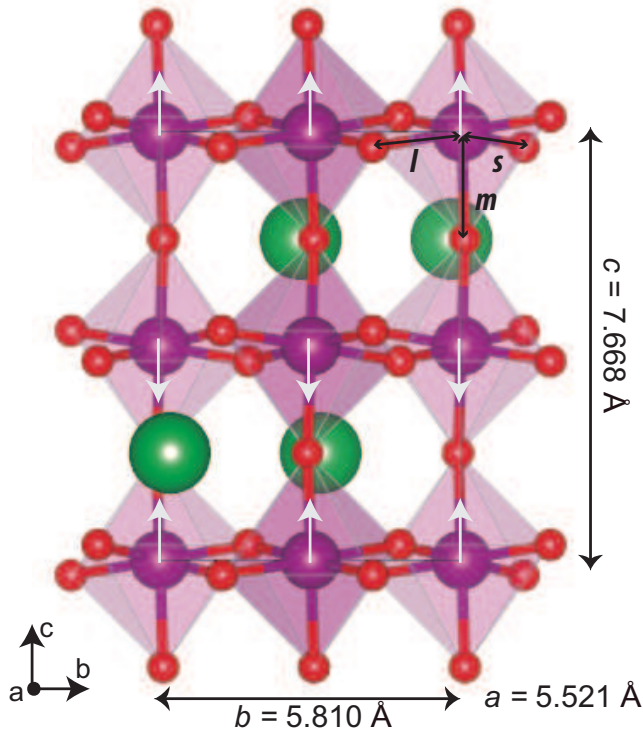}  
\caption{Calculated ground-state Pbnm orthorhombic crystal structure of LaMnO$_3$ showing MnO$_6$ octahedra tilts, rotations, and JT distortion along with the long ($l$), medium ($m$), and short ($s$) Mn-O bonds and A-AFM spin ordering. }  
\label{fig1}  
\end{figure}  

In its ground-state, LaMnO$_3$ forms a Pbnm orthorhombic structure with an A-type antiferromagnetic (A-AFM) insulator character (Fig. 1). The crystal structure contains tilted and rotated MnO$_6$ octahedra along with cooperative JT instabilities characterized by two short ($s$ = 1.903 \r{A}), two long ($l$ = 2.184 \r{A}), and two medium  ($m$ = 1.957 \r{A}) Mn-O bond distances \cite{exp0}. These distortions lift the degeneracy of the singly occupied e$_g$ orbital and stabilize the system into an orbital-ordered state \cite{ord1,ord2}.  Hole-doping or raising the temperature produce a $\textit{pseudo-cubic}$ structure with FM half-metallic character \cite{halfmetallic,our,carvajal} and small volume decrease of 2-4$\%$. 

In contrast, applying 32 GPa of hydrostatic pressure drives the system to a FM metal (but not a half-metal) with a significantly reduced CMR behavior -- in comparison to the hole-doped manganites -- and with an associated 14$\%$ decrease in volume leaving the Pbnm crystal structure unchanged \cite{hydrostatic,raman,franchini2}. DFT calculations have shown that the highly desired half-metallic state can be obtained in the \textit{pseudo-cubic} structure by applying hydrostatic pressures $\ge$ 100 GPa \cite{franchini2}  with an associated drastic reduction in volume of $\simeq$ 35$\%$. As we will show, these large pressures and volume reductions can be significantly reduced by the application of uniaxial pressure while maintaining a FM half-metallic \textit{pseudo-cubic} phase.

In our previous work on LaMnO$_3$ \cite{our}, we established that the Q2 JT mode plays a dominant role in the transition from the low-T AFM insulating state to the high-T FM half-metallic state with the Q3 mode playing a secondary role. In addition, we observe a 2.9$\%$ reduction of the $b$ lattice parameter and an expansion of both $a$ (0.8$\%$) and $c$ (2.8$\%$) lattice parameters that produces a modest $\simeq$ 2.6$\%$ reduction in volume. Combining these two key pieces of information leads to the hypothesis that by applying uniaxial pressure along the $b$ lattice direction ([010]) the pressure needed for obtaining the $\textit{pseudo-cubic}$ FM half-metallic state can be significantly reduced. Specifically, our simulations show there exists strong competition between the FM and AFM magnetic states for a range of uniaxial pressure values that could indicate the emergence of a possible phase coexistence, and hence, the likely realization of CMR at considerably lower pressure. The manifestation of FM half-metallic character with the possibility of a CMR state at relevant technological pressures in the stoichiometric LaMnO$_3$ compound could have a significant impact on spintronic applications.

Unlike the system under high pressure, which lowers its free-energy by creating interfaces (domains, etc.), it is unclear whether such a PS state exists at low pressure. We will address this particular aspect and highlight the possible coexistence of magnetic and structural features that occur for a specific range of uniaxial pressure near the phase transition. The application of compressive uniaxial strain has been investigated for a large variety of materials, with the goal of tuning and optimizing the electronic and magnetic properties \cite{tio2,bicoo,cafeas}, but only a few studies on LaMnO$_3$ have been reported to-date. These studies were focused on providing an interpretation of experiments on epitaxially grown LaMnO$_3$ \cite{hamilt1,hamilt2} and therefore, only applied small uniaxial strains of the order of 2$\%$ or less were considered. It should be mentioned that biaxial in-plane strains have also been applied with the same purpose in a strain range of up to $\pm$ 4$\%$ \cite{epistrain,nanda}. Note that larger strains will not allow high quality epitaxial growth and therefore, have not been studied. Moreover, all reported strain-engineering studies on LaMnO$_3$ were not able to see the possibility of using strain to obtain a half-metallic state. To our knowledge, this is the first time any research has been undertaken to investigate the uniaxial compressive strain and find that $\simeq$ 5$\%$ of compressive strain along the [010] direction is needed to obtain a pseudo-cubic FM half-metallic state in LaMnO$_3$. Furthermore, the small AFM-FM energy differences obtained over 4$\%$ show the emergence of a possible phase coexistence, that along with the half-metallic character, may lead to the realization of a CMR effect.

The possibility of forming a half-metallic state under uniaxial pressure is appealing since, under uniaxial pressure, both the $a$ and $c$ lattice parameters can relax to allow for a much smaller reduction in volume. In contrast, hydrostatic pressure is applied uniformly in all directions and leads to a large volume change. Similarly, one would expect a larger reduction in volume when applying biaxial strain, since the area in the plane where the pressure is applied is reduced, allowing only one direction to relax and/or reconstruct. Weather the PS state can be formed, what type of PS state is obtained,
and if this structure is well-suited for the manifestation of CMR effect are open questions that remain to be addressed. In this work, we identify a range of values for the $b$ lattice parameters where AFM-FM energy differences are significantly small (\textless 20 meV). These small magnetic energy differences suggest the possibility of a phase coexistence that may support a PS state. If this situation can be achieved then it may lead to the realization of a CMR state with optimal transport properties.

\section{Computational and theoretical methods} 

We used the replicated-data version of the \textsc{CRYSTAL14} computational package \cite{CRYSTAL14,CRYSTAL142}. \textsc{CRYSTAL14} is a first-principles electronic structure software package that employs atom-centered Gaussian-type orbital (GTO) basis sets to build the Bloch functions to expand the one-electron crystalline orbitals \cite{dovesi}. GTOs have the advantage of being a local basis set with minimal overlap with neighboring GTO’s. This property provides a high degree of computational performance for calculating the Hartree-Fock exchange (HFX). The numerical cost scales at a rate between \textit{N$^2$-N$^3$}, whereas plane-waves based methods scale as $N^4$, where $N$ is the number of atoms making up the system. This significant improvement in computational performance allows for the simulation of large-scale systems using hybrid DFT functionals. This is an important improvement since the accurate description of the electron-electron correlation in LaMnO$_3$ cannot be achieved with local or semi-local functionals.

A PBE hybrid functional \cite{becke} based on a mixing of 10$\%$ HFX with 90$\%$ PBE exchange potential (PBE-10) has been employed for all calculations. This particular functional has been shown to accurately capture a range of electron-electron correlation effects governing the physics of LaMnO$_3$ systems. This includes the A-AFM insulating ground-state to the high temperature FM half-metallic state \cite{our}. To confirm the broad applicability of this hybrid functional, we first established that it reproduces the IMT observed for an hydrostatic pressure (see Appendix).
 
All-electron GTO’s were used for all atoms comprising the LaMnO$_3$ system. For Mn, the 6-411d41G atomic basis set generated by Towler \textit{et al.} was used \cite{Towler}. This basis includes 1s, 4sp, and 2d contractions. For the O we used the 8-411d basis set constructed by Cor\`a \cite{Cora}. It contains 1s, 3sp and 1d orbital contractions. For La, the 61111sp-631d basis set with 1s, 6sp, and 3d contractions was obtained from the \textsc{CRYSTAL14} repository \cite{webpage} with the addition of two 4f electron shells to better account for the La-O and Mn-O covalency induced by the octahedral rotations found in this system \cite{cov1,cov2}. For our calculations a  8 $\times$ 8 $\times$ 8 Monkhorst-Pack mesh \cite{monkhorst} was utilized to generate a 170 k-points in the irreducible Brillouin zone. Tolerances of 10$^{-7}$ (ITOL1, ITOL2, ITOL3, and ITOL4, using notations from Ref. \onlinecite{CRYSTAL142}) and 10$^{-14}$ (ITOL5) were used for the bielectronic Coulomb and HF exchange series integrals. These settings were found to ensure a convergence of 0.001 eV. The total energy convergence threshold between geometry optimization steps was set to 10$^{-6}$ Ha.

To study the electronic structure and magnetic properties of LaMnO$_3$ under uniaxial pressure along the $b$ axis, we performed constrained geometry relaxation where we fixed the $b$ lattice parameter and relaxed the $a$ and $c$ lattice parameters and all atomic coordinates. Manually changing the $b$ lattice parameter allows for the investigation of different compressive strains applied parallel to the [010] direction. The convergence criterion on gradient components and nuclear displacements was set to 0.0003 Ha/Bohr and 0.0012 Bohr respectively.

\section{III. RESULTS AND DISCUSSION} 
 
We study the competition between different magnetic orderings as a function of $b$ lattice parameter (Subsection A) and provide the structural (Subsection B), magnetic and electronic (Subsection C) properties of LaMnO$_3$ under uniaxial compressive strain along the [010] direction. We demonstrate that at a critical value of $b$ = 5.52 \r{A}, all these properties undergo a simultaneous phase transition due to the strong interplay among electronic, structural, and magnetic interactions. Finally, we quantify the uniaxial pressure needed to achieve this transition (Subsection D).      
 
\subsection{A. Competition between different magnetic phases under uniaxial compressive strain}

To determine the relative stability between all possible magnetic states as a function of compressive strain along [010] direction, we performed simulations on the three possible AFM spin orderings: A-AFM, C-AFM, and G-AFM. Figure 2 displays the energy of these possible AFM configurations and the FM magnetic ordering relative to the A-AFM state as a function of decreasing $b$. FM and A-AFM magnetic phases exhibit the lowest energy for all considered compressive strains between $\varepsilon$$_b$ = 0 and 7$\%$. Within this range of strains, G-AFM and C-AFM stabilize at a much higher energy (over 200 meV) corresponding to a temperature of over 2300 K. Thus, we will omit further analysis of these magnetic phases and the notation A-AFM and AFM will be used interchangeably from this point on. 

We observe that the total energy difference between the AFM and FM states (defined as \textit{$\Delta$E = E$_{AFM}$-E$_{FM}$}) decreases with decreasing $b$. At a critical value of $b=5.52$ \r{A}, a magnetic phase transition from AFM to FM state occurs, which corresponds to a compressive strain of $\varepsilon$$_b$ = 5$\%$ (marked by the symbol "x" on Fig. 2). Additional increases in compression strain further stabilizes the FM phase. The relatively small difference between the AFM and FM states in a narrow range of $b$ values near the transition suggests the possible presence of a particular state characterized by the existence of different magnetic domains. As we will show later, this phase transition is accompanied by an IMT and a significant reduction of the JT distortions, especially the Q2 mode, which is very sensitive to the in-plane lattice parameters. 
  
\begin{figure} 
\includegraphics[width=0.48 \textwidth]{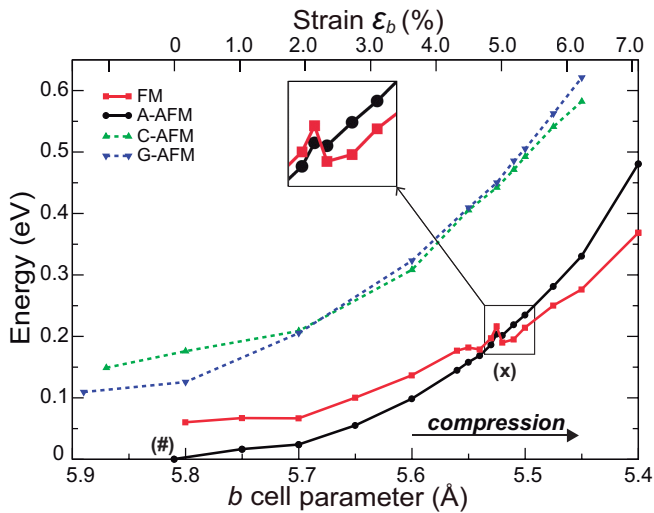} 
\caption{Energy of FM and AFM magnetic orderings relative to A-AFM ground-state energy as function of decreasing $b$ lattice parameter.} 
\label{fig1} 
\end{figure} 

\begin{figure*}[t] 
\includegraphics[width=0.90 \textwidth]{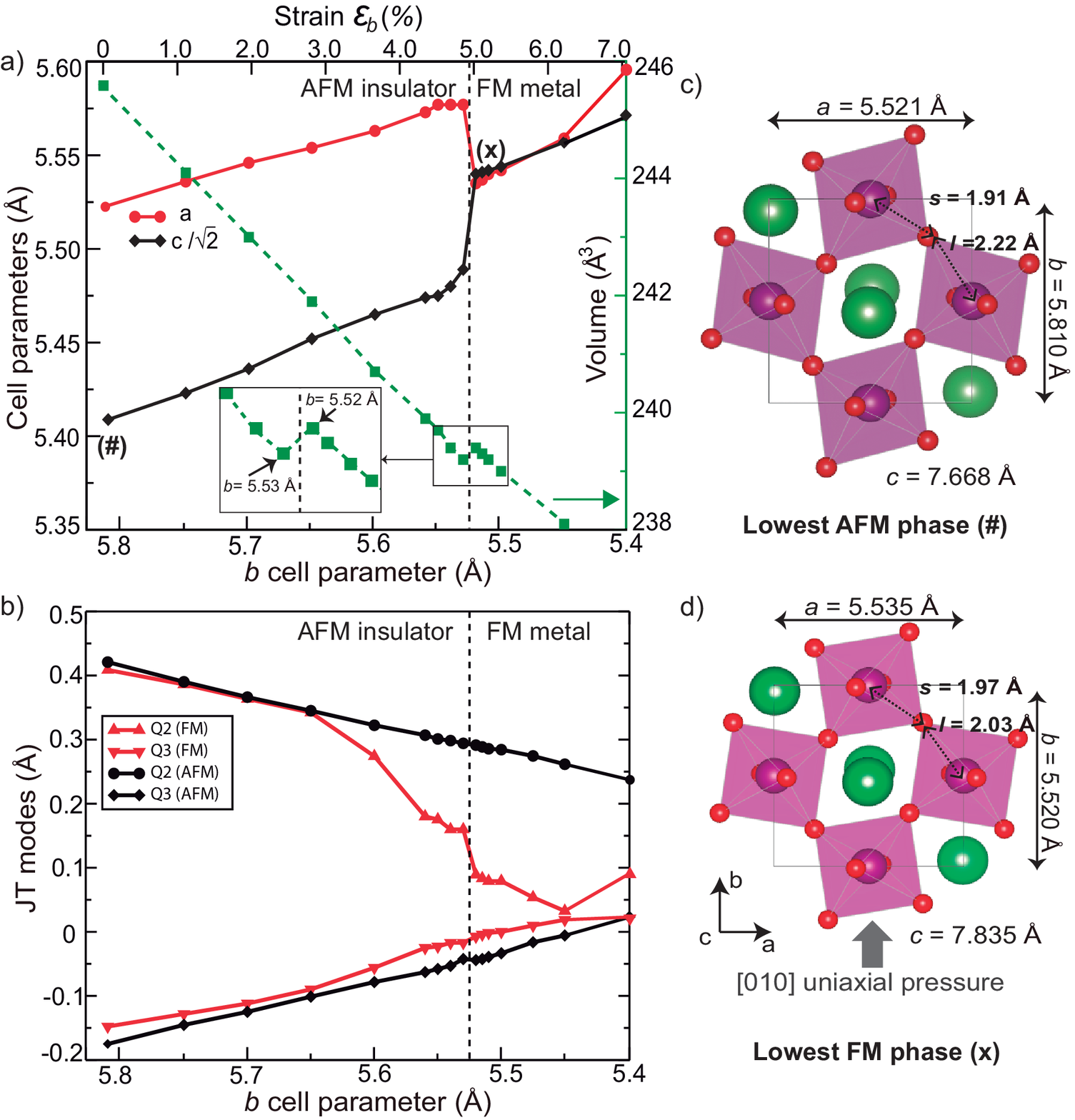} 
\caption{Evolution of a) the structural parameters and volume, and of b) the Q2 and Q3 JT modes as a function of $b$. Q2 and Q3 are defined as $Q2 = 2(l-s)/\sqrt{2}$ and $Q3 = 2(2m-l-s)/\sqrt{6}$.  c) and d) show the top view of the lowest energy structures before (\textbf{\#}) and after the structural transition (\textbf{x}). The inset in a) shows the volume change around the transition. } 
\label{fig3} 
\end{figure*} 

\subsection{B. Structural properties under uniaxial compressive strain}
  
Turning to the structural properties as a function of uniaxial compressive strain, we studied the evolution of the $a$ and $c$ lattice parameters as a function of decreasing $b$ as shown in Fig. 3a. For small values of compressive strain in the range of $\varepsilon$$_b$ = 0 to $\simeq$ 4$\%$ (from $b$ = 5.81 \r{A} to 5.55 \r{A}), we observe a progressive linear expansion of both $a$ and $c$ parameters with $c$ increasing slightly more than $a$. This indicates a slight anisotropic compressive effect on the LaMnO$_3$ structure. When the compressive strain is increased further, small instabilities break the linearity at a value of $\varepsilon$$_b$ $\simeq$ 4-5$\%$. At $b$ = 5.52 \r{A} ($\varepsilon$$_b$ = 5$\%$) the system undergoes a structural phase transition with significant changes in the in-plane lattice parameters. Sharp contraction of $c$ and expansion of $a$ are observed with a corresponding small volume increase. This uniaxial strain produces nearly equal $a$, $b$, and $c$/$\sqrt{2}$ parameters, which indicates the transition from an orthorhombic to a $\textit{pseudo-cubic}$ structure. This structural transformation is accompanied by an AFM insulator to FM half-metal phase transition (see Fig. 4). 

Q2 and Q3 are the two JT distortions that play a major part in the LaMnO$_3$ phase transition (Fig. 3b). Q2 is an in-plane distortion mode while Q3 represents an octahedral stretching mode \cite{vanvleck,kanamori}. These modes are responsible for splitting the Mn 3d-e$_g$ levels, which is necessary for obtaining an AFM insulating state.  While Q3 shows a similar evolution for both AFM and FM states as a function of $b$, Q2 displays marked differences between the two magnetic states. In the AFM phase, Q2 has a nearly linear behavior with a negative slope while the FM phase displays significant non-linear structure with an abrupt change at the structural transition. In other words, the AFM phase contains MnO$_6$ distorted octahedra while the FM phase features nearly regular MnO$_6$ octahedra around the transition. This difference, along with the AFM-FM phase competition observed in Fig. 2 (and later in Fig. 4a) indicates the possibility of developing a phase coexistence state. He \textit{et al.} also observed strong AFM-FM competition in their hydrostatic pressure simulations \cite{franchini2}. However, in contrast to the results presented here, their study predicted that larger and wider range of pressures are needed for this state to be observed. This is not surprising given the strong dependence on the Q2 mode with the application of an uniaxial compressive strain along the $b$-axis, which provides a more direct and energetically favorable path for the phase transition.    
 
\begin{table}[] 
\caption{Calculated $a$ and $c$/$\sqrt{2}$ cell parameters, long ($l$), short ($s$), and medium ($m$) Mn-O bonds, volume ($V$), and $\Delta E$ = $E_{AFM}-E_{FM}$ as a function of compressive strain along the $b$-axis. Units are in \r{A} and meV ($\Delta E$). } 
\label{my-label1} 
\begin{tabular}{lccccccc} 
\hline 
$b$ &$a$&$c/\sqrt{2}$&$l$ &$s$&$m$&$V$&$\Delta E$\\ 
\hline  
5.742 ($^a$) & 5.53 & 5.42 & 2.18 & 1.90 & 1.96 &  243.7 & \textless 0 \\ 
\hline 
5.811 ($^*$)  & 5.52 & 5.41 & 2.22 & 1.91 & 1.96  &  245.2 & -60 \\  
5.75  & 5.54 & 5.42 & 2.19 & 1.92 & 1.96  &  244.2 & -48 \\ 
5.70  & 5.55 & 5.44 & 2.17 & 1.91 & 1.97  & 243.0 & -42 \\  
5.65  & 5.55 & 5.45 & 2.16 & 1.91 & 1.97  &  241.9 & -45 \\  
5.60  & 5.56 & 5.46 & 2.14 & 1.91 & 1.98  &  239.9 & -33 \\  
5.55  & 5.58 & 5.47 & 2.12 & 1.91 & 1.98  &  239.7 & -24 \\  
5.54  & 5.58 & 5.48 & 2.12 & 1.91 & 1.98  &  239.4 & -10 \\  
5.53  & 5.58 & 5.49 & 2.11 & 1.91  & 2.00  &  239.2 & -10 \\  
5.52  & 5.54 & 5.54 & 2.03 & 1.97  & 2.00 & 239.4 & 12 \\ 
5.51  & 5.54 & 5.54 & 2.03 & 1.97  & 2.00 & 239.3 & 19 \\ 
5.50  & 5.54 & 5.54 & 2.02 & 1.97 & 2.00 &  239.0 & 21 \\  
5.45  & 5.56 & 5.56 & 2.00 & 1.98 & 2.00  &  238.1 & 54 \\   
5.40  & 5.59 & 5.56 & 2.02 & 1.96 & 2.00 &  237.4 & 112 \\ \hline 
\end{tabular} 
\\ 
$^a$: Experimental results from Ref.~\onlinecite{exp0} \\ 
$^*$: Fully relaxed\\ 
\end{table} 

The structural effect resulting from the phase transition is to shorten the long Mn-O bond length from $l=2.22$ \r{A} to $l=2.03$ \r{A} and to increase the short and medium bond lengths from $s=1.91$ \r{A} and $m=1.96$ \r{A} to $s=1.97$ \r{A} and $m=2.00$ \r{A} (Figs 3c-d and Table I). This leads to a strong reduction of the Q2 JT distortion. Note, to obtain the FM half-metallic phase under uniaxial compression only requires a 2.3$\%$ volume reduction. This is considerably less than the 14$\%$ volume reduction observed under 32 GPa of hydrostatic pressure needed to produce a similar IMT \cite{raman}, or the $\ge$ 100 GPa to produce a FM half-metallic state \cite{franchini2}. It should be mentioned that data from Fig. 3b and Table I indicate that at $b$ = 5.45\r{A}, both Q2 and Q3 modes are nearly zero with $\Delta E$ = 54 meV (equivalent to $\simeq$ 625 K), suggesting that the system may be in a purely half-metallic FM state.

On the other hand, the bulk phase diagram of LaMnO$_3$ does contain a stable PM phase between the AFM insulating state and the high temperature half-metallic state (that shows FM interactions) with a transition temperature of 750K (65 meV) \cite{FM1,FM2}. Therefore, a clear possibility exists for having a state that contains both PM insulating and FM half-metallic domains. In addition, the range of $b$ values that yields a fully half-metallic FM state is quite small since further reduction of $b$ increases the Q2 mode, thereby increasing the MnO$_6$ distortion that leads to a lowering of the crystal symmetry and eventually to a transition back to the AFM insulating state.    

\begin{figure*}[t] 
\includegraphics[width=0.95 \textwidth]{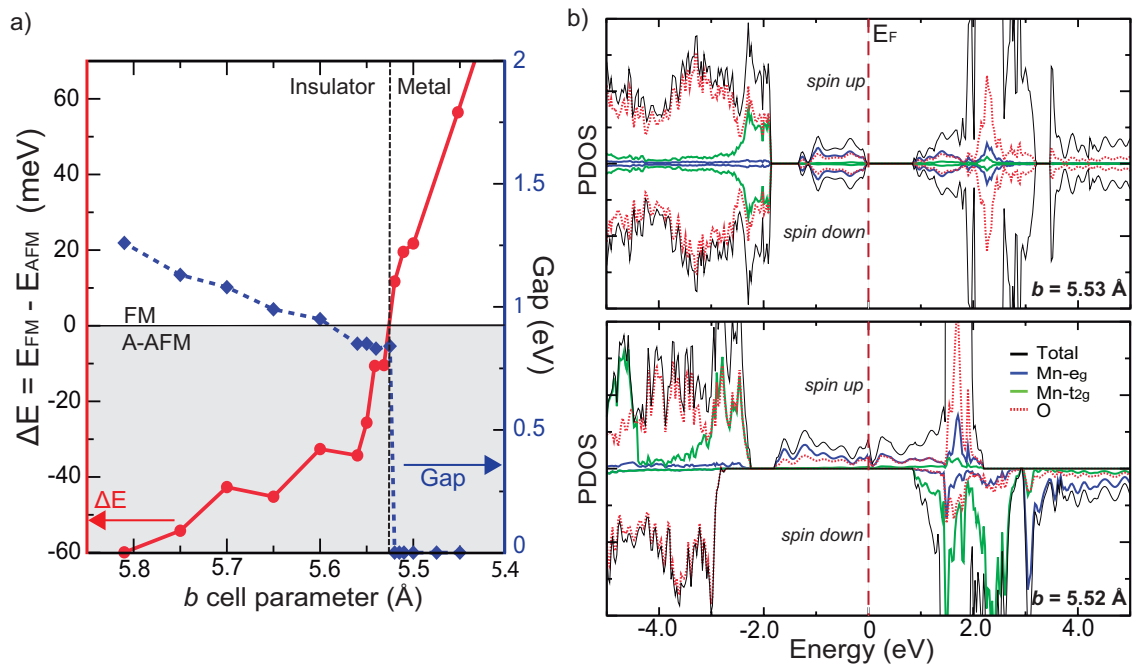} 
\caption{Evolution of the electronic bandgap and total energy difference as a function of $b$. Between $b$ = 5.53 \r{A} and $b$ = 5.52 \r{A}, the systems undergoes an IMT and a magnetic phase transition. b) Spin-up and spin-down projected density of states (PDOS) before and after the transition, showing the transition insulating (top panel) to metallic (bottom panel) phases. } 
\label{fig1} 
\end{figure*} 

 \subsection{C. Electronic and magnetic properties under uniaxial compressive strain}
 
We will now discuss how electronic and magnetic properties evolve when uniaxial compressive strain is applied. Fig. 4a shows the evolution of the total energy difference and of the bandgap as a function of $b$. The bandgap decreases monotonically from 1.26 eV to 0.83 eV for $b$ values decreasing from 5.81 \r{A} to 5.53 \r{A} respectively (compressive strains between 0$\%$ and 5$\%$). This reduction is accompanied by a decrease in $\Delta E = E_{AFM}-E_{FM}$ (from strong AFM coupling to weak AFM). At the critical strain of 5$\%$ ($b$ = 5.52\r{A}), the bandgap drops to 0 indicating an IMT along with magnetic (from AFM to FM) and structural phase transitions occurring simultaneously (Fig. 3). Higher compressive strain applied along the [010] direction further stabilizes the FM state. Furthermore, there exists a range of compressive strains between $b$ = 5.55 \r{A} and $b$ = 5.50 \r{A}, where there is strong competition between AFM and FM phases with corresponding temperatures $\leq$ 230 K. For this range of values, our calculations show that the FM phase is characterized by a $\textit{pseudo-cubic}$ half-metallic state with undistorted MnO$_6$ octahedra, while the AFM phase is represented by an orthorhombic structure with JT distortions (Fig. 3b) and a small bandgap of ~$\simeq$ 0.7 eV. Precisely, this state coincides with the IMT and the emergence of a half-metallic FM phase, which provides the essential ingredients for the manifestation of a CMR state.

Fig. 4b shows LaMnO$_3$'s projected density of states (PDOS) before and after the transition. The Mn 3de$_g$ levels are localized in the conduction and valence bands close to the Fermi level. This is a consequence of the JT distortion and orbital-ordering state of the ground-state structure. The application of uniaxial pressure along the [010] direction initiates a structural phase transformation in LaMnO$_3$ into a $\textit{pseudo-cubic}$ structure. This transition to a higher symmetry structure leads to nearly degenerate e$_g$ states, and therefore provides an additional degree of freedom for electrons, since the probability of occupying either e$_g$ orbital is equal. This results in a FM half-metallic system, in agreement with the predicted transition at high-T \cite{our}. We find some marked differences when we compare this result with the obtained in the IMT driven by hydrostatic pressure \cite{franchini2}. While in the uniaxial case the structure transforms into a $\textit{pseudo-cubic}$ and only the e$_{g}$ states cross the Fermi level, in the hydrostatic case the structure remains orthorhombic and both the e$_g$ and t$_{2g}$ states contribute to the metallic properties.

On the experimental side, recent works have shown the growth of thin films of LaMnO$_{3}$ on SrTiO$_3$ by pulsed laser deposition. SrTiO$_3$ has a cell parameter that would induce the FM half-metallic phase obtained under uniaxial pressure in this work. However, the epitaxial growth of LaMnO$_3$ on SrTiO$_3$ induces a FM insulator behavior \cite{expsubstrate1,expsubstrate2}, probably due to other mechanisms such as the electron transfer at the interface of both materials.

\subsection{D. Quantification of the uniaxial pressure} 
 
We will now establish that both compressive strain and the estimation of the corresponding pressure along the [010] direction can be quantified by an analysis of the stress tensor. The calculated diagonal components of this tensor for the LaMnO$_3$ ground-state structure are found to be less than 0.1 GPa while the off-diagonal terms are less than 10$^{-10}$ GPa, which provides a satisfactory estimation of the initial pressure conditions. The optimization of the atomic positions, and of the $a$ and $c$ cell parameters as a function of decreasing $b$ gives access to the stress tensors needed to quantify the uniaxial pressures applied on the [010] direction. 
 
The diagonal components of the stress tensor $\sigma$$_{xx}$, $\sigma$$_{yy}$, and  $\sigma$$_{zz}$ are shown in Fig. 5 as a function of $b$. As $b$ is decreased, $\sigma$$_{yy}$ increases linearly after the transition while both $\sigma$$_{xx}$ and $\sigma$$_{zz}$ are kept small and nearly constant with corresponding pressures of less than 0.3 GPa for the entire range of $b$ values. The value of the off-diagonal terms are nearly zero for the entire range of compressive strains. The maximum uniaxial stress observed in our calculations occurs before the phase transition at $b$ = 5.53 \r{A}, where $\sigma$$_{yy}$ has a value of 5.40 GPa. Given the relatively small stresses of the $\sigma$$_{xx}$ and $\sigma$$_{zz}$ and the negligible values of the off-diagonal terms, we can estimate that an uniaxial pressure of around 6 GPa is needed to drive the system from the AFM insulator to the FM half-metallic state. 
 
Interestingly, there is a large reduction of more than 3 GPa in $\sigma$$_{yy}$ once the phase transition occurs at $b$ = 5.52 \r{A}. This pressure reduction indicates the structural transition observed in Fig. 3, from the orthorhombic to the $\textit{pseudo-cubic}$ structure. Once this transition is achieved, only $\simeq$ 2 GPa of uniaxial pressure applied along the [010] direction is needed to maintain the FM half-metallic character. Further compressive strain in this direction yields a linear increase of $\sigma$$_{yy}$ while $\sigma$$_{xx}$ and $\sigma$$_{zz}$ remain nearly zero.

\begin{figure}[t] 
\includegraphics[width=0.48\textwidth]{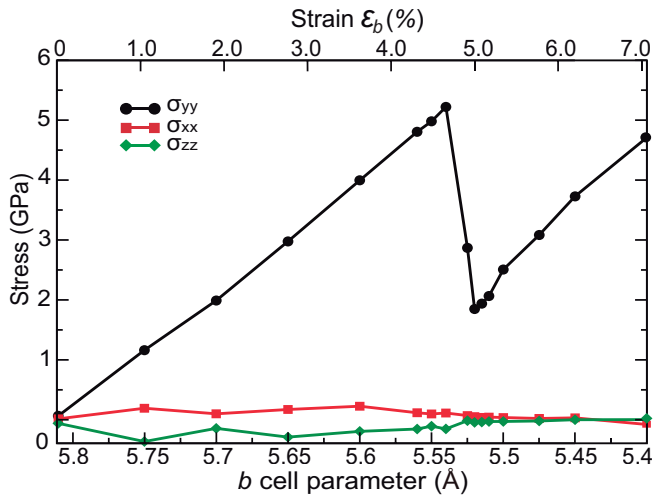} 
\caption{Evolution of the diagonal components of the stress tensor as a function of $b$ cell parameter. Note, the off-diagonal terms remain zero for all the values of $b$ considered.}
\label{fig1} 
\end{figure} 

\section{Conclusions}

We have found that 6 GPa of uniaxial pressure along the [010] direction on LaMnO$_3$ leads to a transition from orthorhombic A-AFM insulator to a $\textit{pseudo-cubic}$ FM half-metallic state with a modest 2.3$\%$ reduction in volume. This pressure is significantly lower than the one previously reported where hydrostatic pressures in excess of 100 GPa are needed to achieve the FM half-metallic state with a very large $\simeq$ 35$\%$ reduction in volume. In addition, the strong competition between FM and AFM magnetic states suggest a possible phase coexistence, that along with the half-metallic behavior may lead to the realization of a CMR effect. Furthermore, the use of uniaxial pressure significantly reduces the 32 GPa of hydrostatic pressure needed to drive an IMT, that leads to a CMR effect but not to the desired half-metallic state. Finally, at $b$ $\simeq$ 5.45\r{A} both Q2 and Q3 JT modes are nearly zero, indicating the possibility of obtaining a half-metal FM undistorted structure that may not be in a phase coexistence state.

The computational work conducted by W.A.S. and P.R is supported by the U.S. Department of Energy under EPSCoR Grant No. DE-SC0012432 with additional support from the Louisiana Board of Regents and by an allocation of computing time from the Louisiana State University High Performance Computing center. V. M. acknowledges support by New York State under NYSTAR program C080117. 

\section{Appendix}

Table II shows the calculated cell parameters, energy difference and bandgaps as a function of hydrostatic pressure via PBE-10. This functional reproduces the low pressure AFM insulator to the high pressure FM metal observed experimentally at 32 GPa \cite{raman} and predicts this transition at around $\simeq$ 22 GPa, far below the result obtained using HSE-15 ($\simeq$ 50 GPa) \cite{franchini2}. 
Unlike the uniaxial case, the hydrostatic pressure leads to a orthorhombic metallic state that is not half-metallic.

\begin{table}[]
\centering
\caption{Calculated $a$, $b$, and $c$$\sqrt{2}$ cell parameters (\r{A}), $\Delta E$ = $E_{AFM}-E_{FM}$ (meV), and bandgaps (eV) as a function of hydrostatic pressure (GPa). }
\label{my-label}
\begin{tabular}{cccccc}
P  & $\Delta$E  & a    & b    & c/$\sqrt{2}$    & Gap  \\ \hline
30 & 61  & 5.34 & 5.28 & 5.31 & metallic   \\
25 & 63  & 5.36 & 5.31 & 5.34  & metallic     \\
23 & 8 & 5.37 & 5.33 & 5.36 & metallic      \\
21 & -21 & 5.41 & 5.41 & 5.29 & 0.64 \\
18 & -38 & 5.43 & 5.43 & 5.31 & 0.69 \\
15 & -50 & 5.45  & 5.47 & 5.33 & 0.71 \\ \hline
\end{tabular}
\end{table}

\end{document}